\documentclass[conference]{IEEEtran}
\IEEEoverridecommandlockouts

\usepackage[colorlinks=true,linkcolor=black,anchorcolor=black,citecolor=black,filecolor=black,menucolor=black,runcolor=black,urlcolor=black]{hyperref}
\usepackage{amssymb}
\usepackage{multirow}
\usepackage{subcaption}
\usepackage{caption}
\usepackage{url}
\usepackage{paralist}
\usepackage{cite}
\usepackage{amsmath,amssymb,amsfonts}
\usepackage{algorithmic}
\usepackage{graphicx}
\usepackage{textcomp}
\usepackage{xcolor}
\def\BibTeX{{\rm B\kern-.05em{\sc i\kern-.025em b}\kern-.08em
    T\kern-.1667em\lower.7ex\hbox{E}\kern-.125emX}}

\begin{document}

\title{A privacy-preserving method using secret key \\for convolutional neural network-based \\speech classification\\
}

\author{\IEEEauthorblockN{Shoko Niwa}
\IEEEauthorblockA{\textit{Tokyo Metropolitan University}\\
Tokyo, Japan \\
niwa-shoko@ed.tmu.ac.jp}
\and
\IEEEauthorblockN{Sayaka Shiota}
\IEEEauthorblockA{\textit{Tokyo Metropolitan University}\\
Tokyo, Japan \\
sayaka@tmu.ac.jp}
\and
\IEEEauthorblockN{Hitoshi Kiya}
\IEEEauthorblockA{\textit{Tokyo Metropolitan University}\\
Tokyo, Japan \\
kiya@tmu.ac.jp}
}

\maketitle

\begin{abstract}
In this paper, we propose a privacy-preserving method with a secret key for convolutional neural network (CNN)-based speech classification tasks.
Recently, 
many methods related to privacy preservation have been developed in image classification research fields. 
In contrast, in speech classification research fields, little research has considered these risks. 
To promote research on privacy preservation for speech classification, we provide an encryption method with a secret key in CNN-based speech classification systems. 
The encryption method is based on a random matrix with an invertible inverse. 
The encrypted speech data with a correct key can be accepted by a model with an encrypted kernel generated using an inverse matrix of a random matrix.
Whereas the encrypted speech data is strongly distorted, the classification tasks can be correctly performed when a correct key is provided. 
Additionally, in this paper, we evaluate the difficulty of reconstructing the original information from the encrypted spectrograms and waveforms. 
In our experiments, the proposed encryption methods are performed in automatic speech recognition~(ASR) and automatic speaker verification~(ASV) tasks.
The results show that the encrypted data can be used completely the same as the original data when a correct secret key is provided in the transformer-based ASR and x-vector-based ASV with self-supervised front-end systems. 
The robustness of the encrypted data against reconstruction attacks is also illustrated.

\end{abstract}


\section{Introduction}\label{sec:intro}
In recent years, cloud services have been increasingly used in many applications.
Cloud services have the advantages of reducing initial computer investment and maintenance costs, and facilitating information sharing.
However, since cloud services are managed by external providers, various threats such as data leakage due to malicious attacks from outside or inside are a concern~\cite{tabrizchi2020survey}.
When using classification models on a cloud service, it is necessary to provide a trained model and query data to the cloud service. 
Therefore, when cloud services are insecure, models and queries face threats. 
To prevent such risks, it is important to preserve privacy before sending data to insecure services.

Speech data usually includes personal information such as age, gender, language, and speaking content.
Therefore, the issue of privacy also has been gradually gaining attention as the latest topic in the research field of speech processing~\cite{tomashenko2022voiceprivacy}.
In the research field of image processing, many privacy-preserving methods have been proposed for CNN-based systems~\cite{kiya2022overview, maungprivacy}.
Some latest speech classification systems also adopt a convolutional layer for accepting speech data.
Thus, the privacy-preserving methods for image classification can be easily applied to such CNN-based speech classification systems.

There are two patterns for inputting speech data into a convolutional layer: using a two-dimensional representation such as a spectrogram, and directly using a waveform.
Therefore, we propose privacy-preserving methods that use a secret key to encrypt both spectrograms and waveforms and are assumed to have a random matrix with an invertible inverse.
As examples of encryption methods with random matrices with invertible inverses, we propose two methods: \textit{shuffling} and \textit{flipping}. 
Speech data encrypted by either encryption method is highly distorted compared with its original data. 
Therefore, the classification task can only be performed correctly when a correct key is provided.
In the experiments, we performed the proposed privacy-preserving methods in automatic speech recognition (ASR) and automatic speaker verification (ASV).
From the results, we confirmed that when a correct secret key is used, the classification performances are completely the same as those without encryption, and when an incorrect key is used, the accuracies are significantly decreased.
Additionally, in this study, we evaluate the difficulty of reconstructing the original information from the encrypted spectrograms and waveforms. 
Regarding sound reconstruction methods, phase reconstruction approaches and decryption attacks can be considered to reconstruct the original waveforms from spectrograms~\cite{pghi2017}. From these reconstruction experiments, the proposed methods can be shown to have high-security performance.
To evaluate the difficulty of reconstructing the original spectrograms from the encrypted spectrograms and waveform, phase reconstruction and a decryption attack were performed on encrypted speech data.

In the following, we outline the structure of the paper.
In Section~\ref{sec:image}, we describe the privacy-preserving classification scenario. 
In Section~\ref{sec:proposed}, we describe the details of the proposed method and the attack issue of the proposed method, and in Section~\ref{seq:experiment} we show the results. 
In Section~\ref{seq:conclude}, we conclude the study and describe our future work.

\begin{figure}[t]
    \centering
    \includegraphics[keepaspectratio,width=7.9cm]{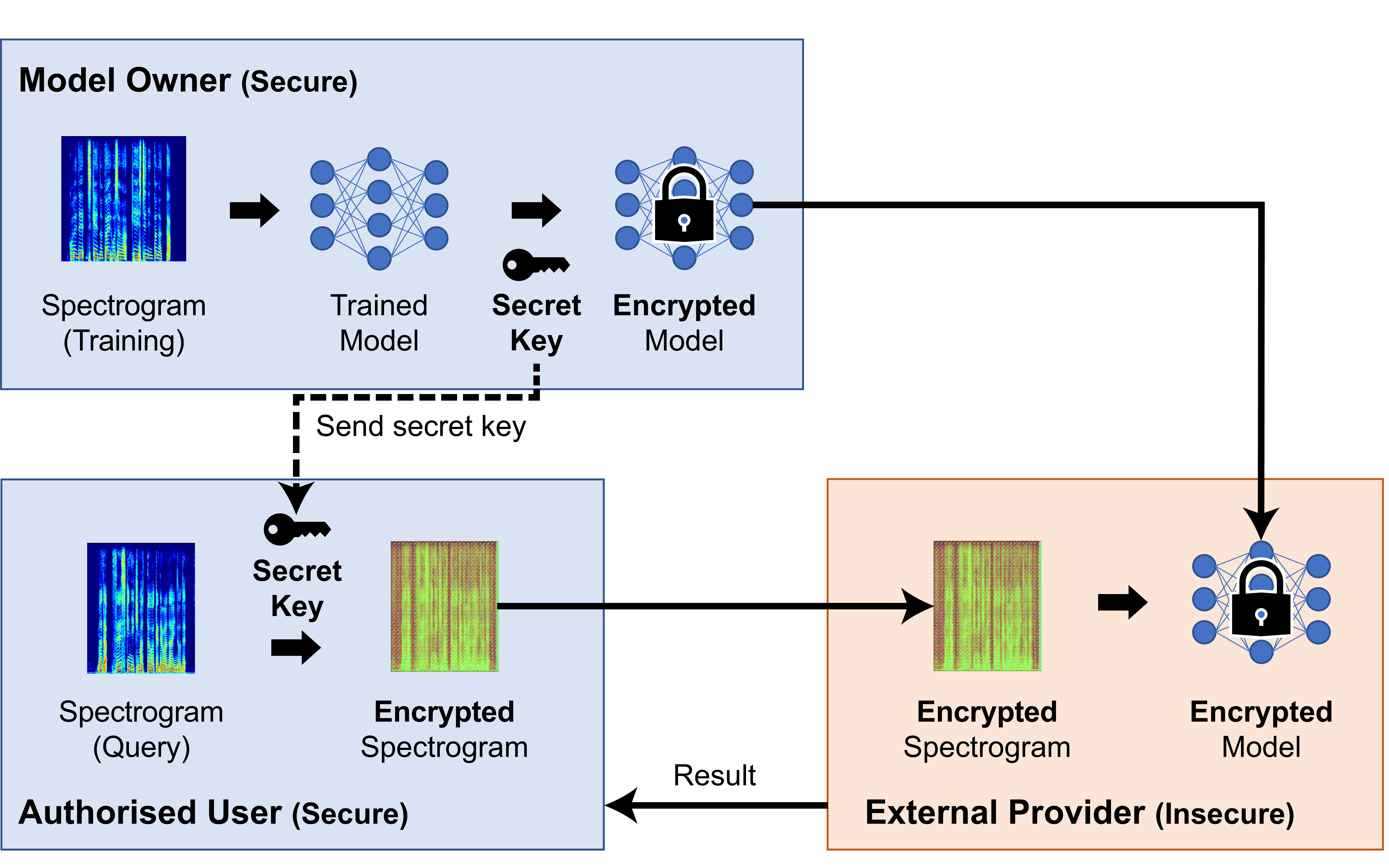}
    \caption{Privacy-preserving scenario}
    \label{fig:gaiyou}
\end{figure}

\section{Privacy-preserving classification scenario}\label{sec:image}
Generally, there are three types of privacy-preserving issue for machine learning-based systems:
\begin{inparaenum}[(1)]
\item privacy of datasets,
\item privacy of models, and
\item privacy of models' outputs~\cite{privacypreserving}.
\end{inparaenum}
In this paper, we focus on speech classification considering privacy preservation in terms of the privacy of datasets.
The privacy-preserving scenario is illustrated in Fig.~\ref{fig:gaiyou}.
It is based on privacy preservation in image classification~\cite{kiya2023image}.
From Fig.~\ref{fig:gaiyou}, first, a model owner trains a classification model with plain speech data, e.g., spectrograms and waveforms in a secure environment. 
Then, the trained model is encrypted with a secret key. 
Since the encryption is performed after training, the secret key can be changed easily without retraining the model.
Next, the model owner provides the encrypted model to an external provider, such as a cloud service, and shares the secret key with an authorized user.
When the authorized user wants to use the encrypted classification system published by the external provider, an encrypted query with the shared secret key is sent.
In this scenario, only an authorized user who knows the correct key can use the encrypted model.
In comparison, an unauthorized user who does not know the correct key cannot use the model correctly and cannot reconstruct a speech utterance from the encrypted query.
In this framework, it is assumed that the environment of model owners and authorized users is secure and that of external providers is insecure.
Since an external provider performs classification by using the encrypted queries and models, the privacy information in spectrograms is protected even if the external provider is insecure.

\section{Proposed methods}\label{sec:proposed}

\subsection{Query encryption}\label{queryenc}
In this subsection, we describe a speech data encryption method that use a secret key.
Basically, the procedure is followed to~\cite{iijima2022encryption}. 
For adapting to speech data such as two-dimensional spectrograms, speech data is encrypted by whichever encryption method can be obtained through the following procedure. 
First, let speech data~$X$ be regarded as a spectrogram; speech data~$X$ with a size of $T\times F$ is divided into blocks with a size of $M\times M$ each, where $T$ is the size of $X$ in the time direction, $F$ is the size of $X$ in the frequency direction and $M$ is the block size.
Next, $b$-th block~$X_\text{b}$ is flattened into a one-dimensional vector with a size of $M^2$. 
Then, $X_\text{b}$ is converted into $X'_\text{b}$ by an encryption method while maintaining the dimension using a secret key.
There are two methods for converting $X_\text{b}$ to $X'_\text{b}$ as follows;

\subsubsection{Shuffling}
A secret key~$K_\text{s}$ is an array of randomly permuted indices whose size is $M^2$, and its key space refers to the number of possible keys, denoted as $M^2!$.
$X_\text{b}$ is transformed into $X'_\text{b}$ using $K_\text{s}$ as follows:
\begin{equation}\label{eq:pix}
X'_\text{b}(i) = X_\text{b}(K_\text{s}(i)) \text{,}
\end{equation}
where $1\leq i \leq M^2$.
Eq.~\eqref{eq:pix} shuffles the positions of the spectrogram values in $X_\text{b}$ according to $K_\text{s}$.
When the speech data~$X$ is regarded as a waveform, $F_\text{b}$ is set to one in the procedure.
In this case, the key space of $K_\text{s}$ is $M!$, and the positions of the values included in the waveform can be shuffled by Eq.~\eqref{eq:pix}.

\subsubsection{Flipping}
A secret key~$K_\text{f}$ is a bit sequence, and zero and one are generated with equal probability.
The size of $K_\text{f}$ is $M^2$ and its key space is $2^{M^2}$. 
$X_\text{b}$ is transformed to $X'_\text{b}$ using $K_\text{f}$ as follows:
\begin{equation}\label{eq:bit}
X'_\text{b}(i) = 
\begin{cases}
-X_\text{b}(i)&(K_\text{f}(i)=1)\\
X_\text{b}(i)&(K_\text{f}(i)=0)
\end{cases} \text{,}
\end{equation}
where $1\leq i \leq M^2$.
Eq. \eqref{eq:bit} inverts the sign of the spectrogram values in $X_\text{b}$ according to $K_\text{f}$.
When $X$ is a waveform, the key space of $K_\text{f}$ is $2^{M}$ and the transformation is performed according to Eq.~\eqref{eq:bit}.

Finally, each encrypted blocks~$X'_\text{b}$ is integrated to obtain the encrypted spectrogram~$X'$ with a size of $T\times F$.

\begin{figure}[t]
\centering
  \begin{minipage}[b]{0.4\linewidth}
    \centering
    \includegraphics[keepaspectratio, width=3.2cm]{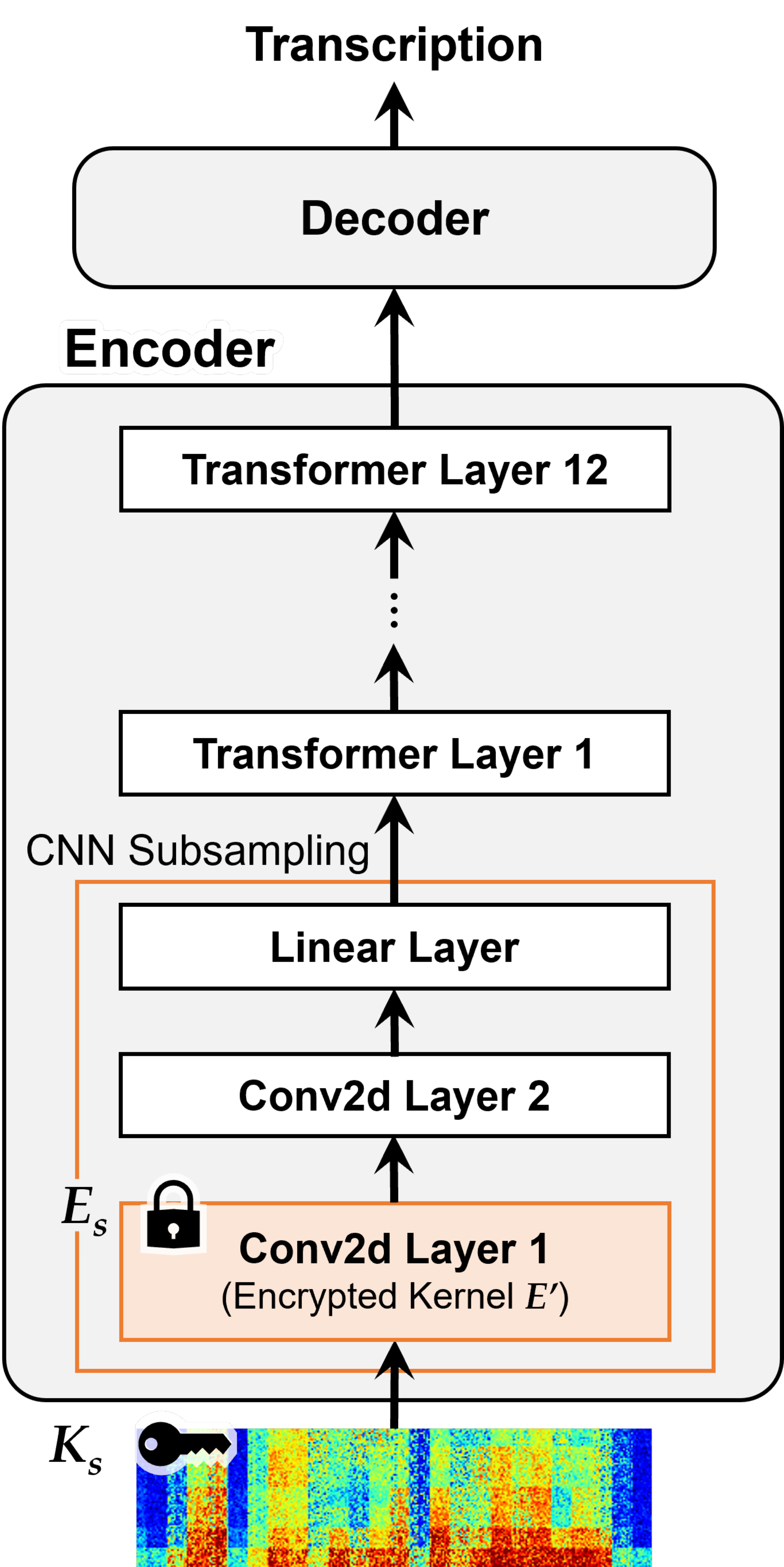}
    \subcaption{ASR system~\cite{ESPnetASR}}
    \label{fig:asr}
  \end{minipage}
  \hspace{0.05\columnwidth}
  \begin{minipage}[b]{0.4\linewidth}
    \centering
    \includegraphics[keepaspectratio,width=3.2cm]{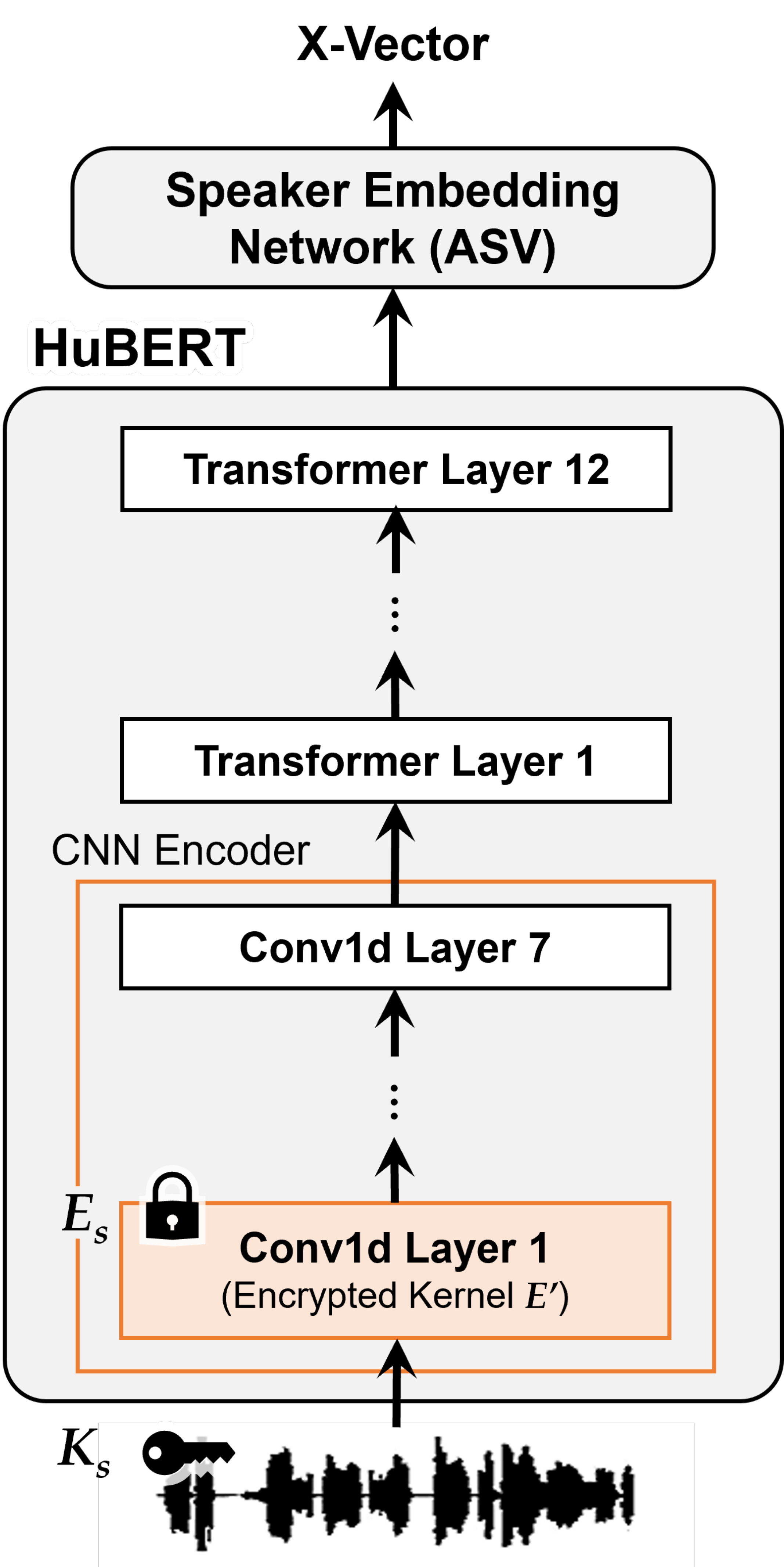}
    \subcaption{ASV system~\cite{hubert}}
    \label{fig:asv}
  \end{minipage}
  \caption{Examples of models accepting encrypted speech data}
  \label{figs:model-enc}
\end{figure}

\subsection{Models accepting encrypted queries}
To accept the encrypted queries with a secret key, the parameters of the trained classification model are transformed.
The proposed encryption methods assume a classification model including a convolutional layer for accepting input data.
When the kernel size and stride size of the convolutional layer are equal, it means that the input data is divided into patches without overlap.
Since each encrypted block $X'_\text{b}$ is treated independently, 
in this paper, the patch size is equal to the kernel and stride sizes. 
Let us denote the patch size of the convolutional layer to be encrypted as $P$, the number of output dimensions as $d$, and the block size as equal to the patch size; then, the kernel of the patch embedding layer can be expressed as $E\in \mathbb{R}^{P\times P\times d}$.
To cancel out each encryption, the kernel of the model is defined as follows:

\subsubsection{Shuffling}
Kernel~$E$ is permuted with a shared secret key~$K_\text{s}$ so that it can correctly accept the encrypted data into a classification model.
The encrypted kernel~$E'$ is defined using the permutation matrix~$E_\text{s}$, which is defined with $K_\text{s}$, as follows:
\begin{equation}
E' = E_\text{s} E \text{.}
\end{equation}
We can treat encrypted data~$X'$ without retraining by simply attaching the encrypted kernel~$E'$ to the first convolutional layer.

\subsubsection{Flipping}
Inverting the sign of the weights contained in kernel~$E$ with a shared secret key~$K_\text{f}$ is necessary to correctly accept the encrypted data into the encrypted model.
The transformation from kernel~$E$ into encrypted kernel~$E'$ is shown as follows:
\begin{equation}
E'[i,j] =
\begin{cases}
-E[i,j]&(k_l=1)\\
E[i,j]&(k_l=0)
\end{cases}\text{,}
\end{equation}
where $l = (i-1)P+j, 1\leq i,j\leq P$ and $k_l$ is the $l$-th element of $K_\text{f}$.

Figures~\ref{fig:asr} and~\ref{fig:asv} are examples of ASR and ASV models that were converted to accept data encrypted by shuffling. In this paper, we performed transformations on the first convolutional layer of each model.
In the case of flipping, encrypted data~$X'$ can also be inputted without retraining by attaching the encrypted kernel~$E'$ in the same flow as in Fig.~\ref{figs:model-enc}.

\subsection{Attacks on encrypted data}
To evaluate the security of encryption methods, it is necessary to demonstrate the difficulty of recovering the original data by performing decryption attacks.
In the image processing research field, decryption attacks are performed on encrypted images to restore visual information on plain images~\cite{maungprivacy}.
In the speech processing research field, decryption attacks are performed on encrypted spectrograms or waveforms to restore the original data.

In this paper, to evaluate the privacy-preserving performance, we applied a phase reconstruction method and decryption attacks.
The phase reconstruction method developed by Průša \textit{et al.}~\cite{pghi2017}, was adopted on the encrypted spectrograms to investigate whether the waveforms before encryption could be reconstructed.
Since there are few attacks against encrypted spectrograms, we adopted Alex \textit{et al.}'s~\cite{attack} method of the decryption attack on encrypted images. The encrypted spectrograms are attacked as encrypted images.

\begin{table}[t]
\caption{WER$(\%)$ in the encryption scenario ($M=3$) for ASR on LibriSpeech~\cite{libri} (test\_clean / test\_other subsets).}
\label{tab:asr-result}
\centering
\begin{tabular}{c|c|cc}
\hline
\begin{tabular}{c}\vspace{-0.5mm}
\end{tabular}
& Plain & 
\begin{tabular}{c}\vspace{-1mm}
Correct\\key
\end{tabular}
& 
\begin{tabular}{c}\vspace{-1mm}
Incorrect\\key
\end{tabular}\\
\hline\hline
No encryption (Plain) & 4.4 / 10.5 & - & -\\ 
Shuffling & 12.2 / 24.8 & 4.4 / 10.5 & 11.9 / 24.7\\
Flipping & 97.8 / 98.4 & 4.4 / 10.5 & 97.8 / 98.1\\
\hline
\end{tabular}
\end{table}

\begin{table}[t]
\caption{EER$(\%)$ in the encryption scenario ($M=10$) for ASV on VoxCeleb1 test set~\cite{voxceleb}.}
\label{tab:asv-result}
\centering
\begin{tabular}{c|c|cc}
\hline
\begin{tabular}{c}\vspace{-0.5mm}
\end{tabular}
& Plain & 
\begin{tabular}{c}\vspace{-1mm}
Correct\\key
\end{tabular}
& 
\begin{tabular}{c}\vspace{-1mm}
Incorrect\\key
\end{tabular}\\
\hline\hline
No encryption (Plain) & 8.3 & - & -\\ 
Shuffling & 41.3 & 8.3 & 37.6 \\
Flipping & 39.3 & 8.3 & 39.1 \\
\hline
\end{tabular}
\end{table}

\section{Experiment}\label{seq:experiment}
We evaluated the proposed privacy-preserving methods using ASR and ASV tasks.

\subsection{Experimental conditions}
For the ASR task, we trained a transformer model with the LibriSpeech corpus~\cite{libri} following the ESPnet2 recipe~\cite{Watanabe2018}.
The transformer architecture and hyperparameters were the same as in~\cite{ESPnetASR}, except for the input feature and the stride size of the first convolutional layer.
The input feature was set to 80-dim log-mel filterbank frames.
The stride size of the first convolutional layer, which included the encrypted kernel $E'$, was set to three in order to adopt the proposed method.
The block size~$M$ for the encryption was set to three to match the kernel size of the first convolutional layer.
Word error rate (WER) was used as an evaluation metric.

For the ASV task, we adopted an x-vector-based ASV system~\cite{xvector} with a self-supervised front-end model. 
A HuBERT model~\cite{hubert} trained with the LibriSpeech corpus was used as the self-supervised front-end model.
The structure and hyperparameters of the HuBERT model were the same as those of the HuBERT \textsc{Base}~\cite{hubert}, except that the stride size of the first convolutional layer was changed to $10$.
The block size~$M$ for the encryption was set to $10$.
The speech expression outputted from the HuBERT model was inputted to the x-vector-based embedding network. 
The x-vector-based embedding network was trained with the VoxCeleb1 corpus~\cite{voxceleb}, using the same hyperparameters as in~\cite{superb}. 
Equal error rate (EER) was used as the evaluation metric.

We performed the proposed method in three scenarios: ``Correct key'', ``Incorrect key'', and ``Plain''. 
``Correct key'' means that both keys used for encrypting the model and the queries were the same, ``Incorrect key'' means that both keys were not matched, and ``Plain'' means that only the model was encrypted, and the query was not encrypted.

\begin{figure}[t]
\centering
  \begin{minipage}[b]{0.3\linewidth}
    \centering
    \includegraphics[keepaspectratio,width=2.3cm]{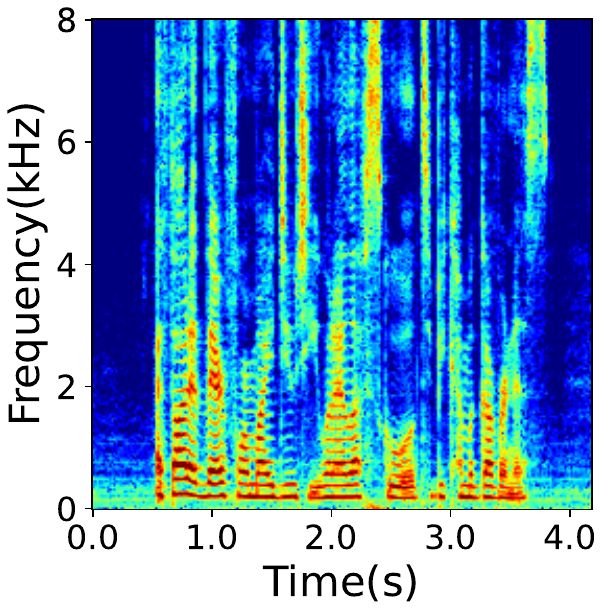}
    \subcaption{Original\\(Plain)}
    \label{fig:ori1}
  \end{minipage}
  \hspace{0.02\columnwidth}
  \begin{minipage}[b]{0.3\linewidth}
    \centering
    \includegraphics[keepaspectratio, width=2.3cm]{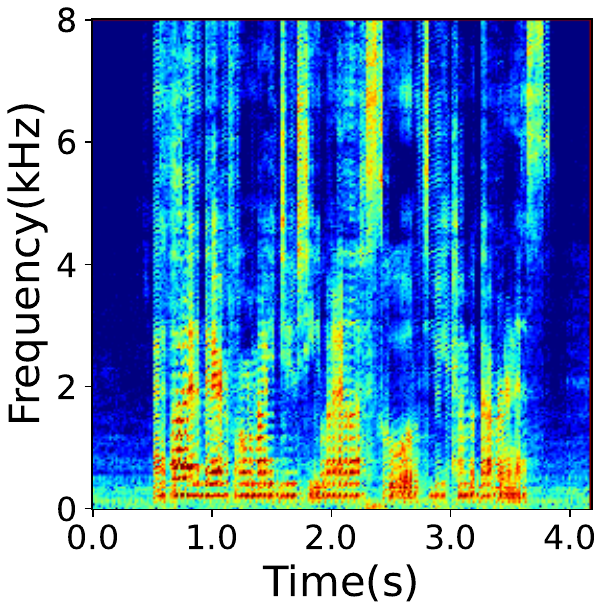}
    \subcaption{Shuffling \protect \linebreak $(M=3)$}
    \label{fig:pixel_shuffling3}
  \end{minipage}
  \hspace{0.02\columnwidth}
  \begin{minipage}[b]{0.3\linewidth}
    \centering
    \includegraphics[keepaspectratio, width=2.3cm]{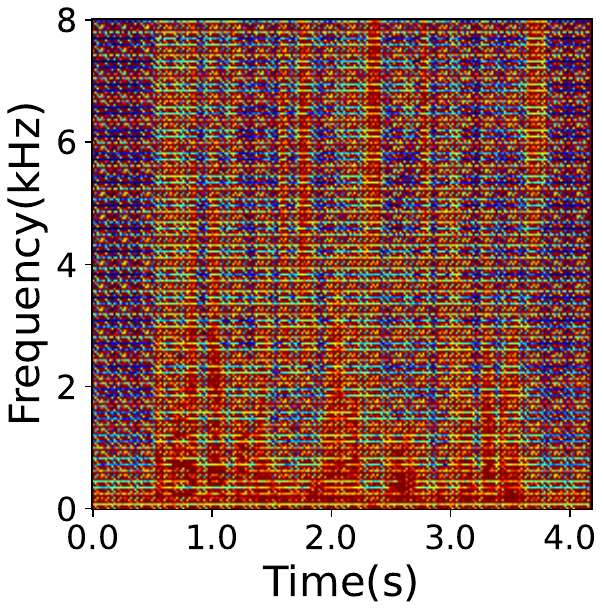}
    \subcaption{Flipping $(M=3)$}
    \label{fig:bit_flipping3}
  \end{minipage}
  \caption{Spectrograms encrypted with proposed method.}
  \label{figs:ps-bf}
\end{figure}

\subsection{Experimental results}
\subsubsection{ASR task}
Table~\ref{tab:asr-result} shows the results of the ASR experiments using the no-encryption (plain) model and the models encrypted by shuffling and flipping.
The WERs of the encrypted models in Correct key were completely the same as those of the no-encryption model.
This indicates that the Correct key scenario was performed as we expected.
In the Incorrect key scenario, the WERs were higher than those of the Correct key scenario.
In particular, when flipping was applied as the encryption method, the WERs increased significantly.
To analyze these results, the spectrograms encrypted by the proposed method are shown in  Fig.~\ref{figs:ps-bf}.
Figure~\ref{fig:ori1} shows the original plain spectrogram, and Figs.~\ref{fig:pixel_shuffling3} and~\ref{fig:bit_flipping3} show the spectrograms encrypted by shuffling and flipping, respectively, under the condition $M=3$.
By comparing Figs.~\ref{fig:ori1} and~\ref{fig:pixel_shuffling3}, we can see that the positions of values in each block move in accordance with the secret key, and the harmonic structure of the spectrogram is distorted.
By comparing Figs.~\ref{fig:ori1} and~\ref{fig:bit_flipping3}, we can confirm that the magnitude of each value in the encrypted spectrogram changes randomly.
The value in the spectrogram greatly increases owing to the sign inversion, especially in the silence intervals in Fig.~\ref{fig:ori1}, so the encrypted spectrogram is markedly different from the original one.
A larger block size~$M$ results in a larger key space and better privacy-preserving performance.
On the other hand, it will affect the accuracy of ASR in the Plain scenario, so there is a trade-off relationship between accuracy and privacy-preserving performance.

\subsubsection{ASV task}
\begin{figure}[t]
\centering
  \begin{minipage}[b]{0.32\linewidth}
    \centering
    \includegraphics[keepaspectratio,height=1.9cm]{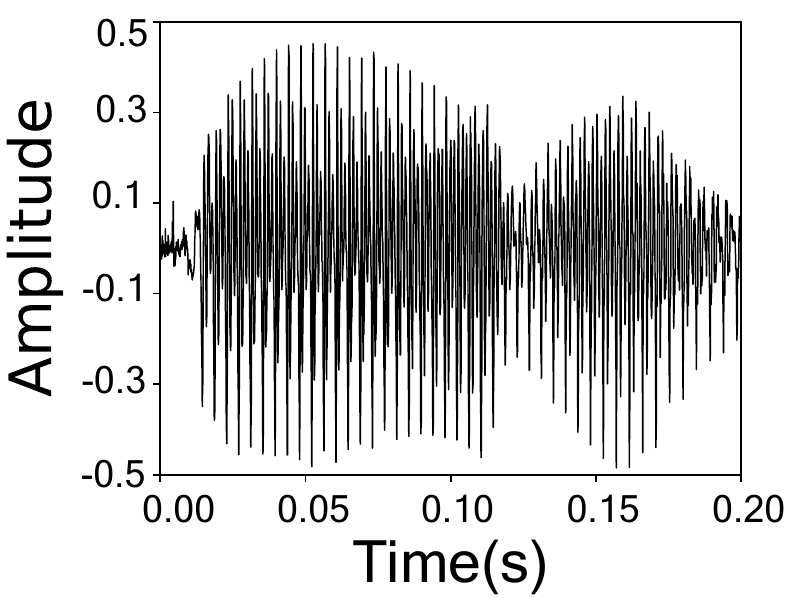}
    \subcaption{Original}
    \vspace{0.8\baselineskip}
    \label{fig:ori-wav}
  \end{minipage}
  \begin{minipage}[b]{0.32\linewidth}
    \centering
    \includegraphics[keepaspectratio,height=1.9cm]{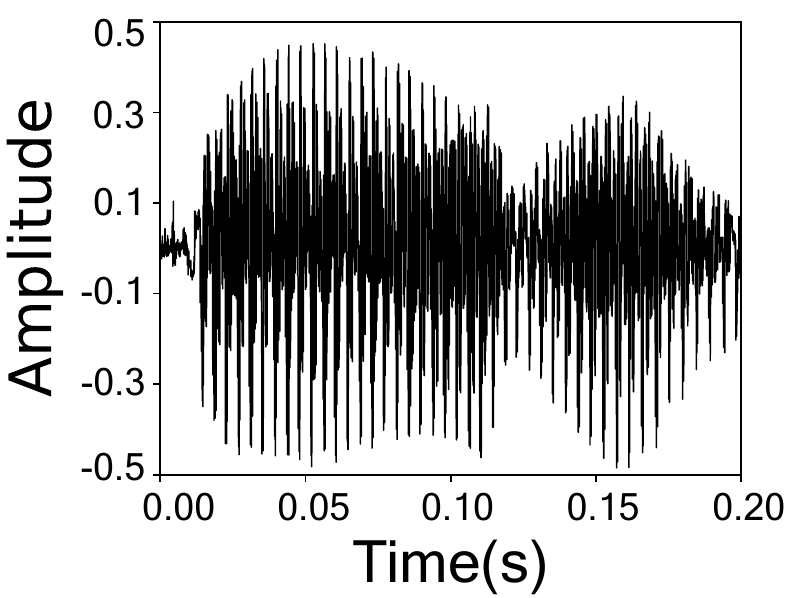}
    \subcaption{Shuffling ($M=10$)}
    \label{fig:pix32-wav}
  \end{minipage}
  \begin{minipage}[b]{0.32\columnwidth}
    \centering
    \includegraphics[keepaspectratio,height=1.9cm]{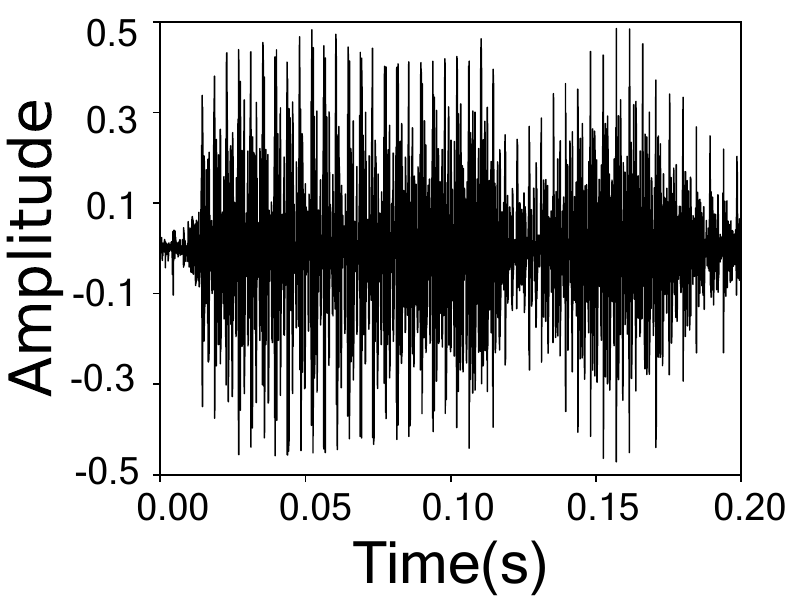}
    \subcaption{Flipping ($M=10$)}
    \label{fig:bit32-wav}
  \end{minipage}
  \caption{Waveforms encrypted by proposed method.}
  \label{figs:wav-enc}
\end{figure}

\begin{figure}[t]
\centering
  \begin{minipage}[b]{0.3\linewidth}
    \centering
    \includegraphics[keepaspectratio,width=2.3cm]{fig/ori-spe.pdf}
    \subcaption{Original}
    \vspace{0.8\baselineskip}
    \label{fig:ori-spe}
  \end{minipage}
  \hspace{0.01\columnwidth}
  \begin{minipage}[b]{0.3\linewidth}
    \centering
    \includegraphics[keepaspectratio, width=2.3cm]{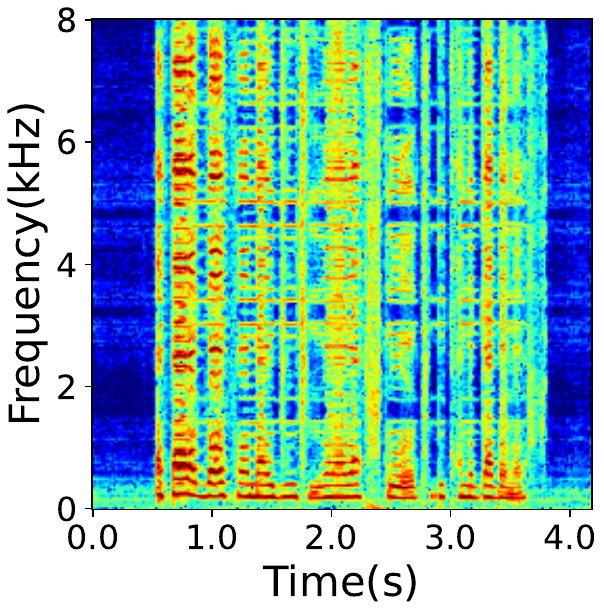}
    \subcaption{Shuffling ($M=10$)}
    \label{fig:pix32-spe}
  \end{minipage}
  \hspace{0.01\columnwidth}
  \begin{minipage}[b]{0.3\columnwidth}
    \centering
    \includegraphics[keepaspectratio, width=2.3cm]{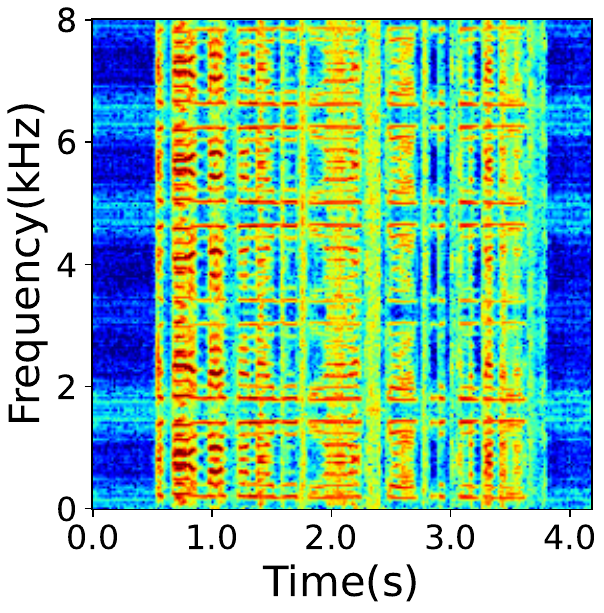}
    \subcaption{Flipping ($M=10$)}
    \label{fig:bit32-spe}
  \end{minipage}
  \caption{Spectrograms corresponding to Fig.~\ref{figs:wav-enc}.}
  \label{figs:spe-wav-enc}
\end{figure}

Table~\ref{tab:asv-result} shows the results of the ASV experiments using the no-encryption (plain) model and the models encrypted by shuffling and flipping.
Similarly to the ASR results, the EERs of the Correct key were completely the same as those of the no-encryption model.
In the Incorrect key scenario, the EERs were higher than those of the the Correct key scenario.
To analyze these results, the waveforms encrypted by the proposed method are shown in  Fig.~\ref{figs:wav-enc}.
Figure~\ref{fig:ori-wav} shows the original plain waveform, Figs.~\ref{fig:pix32-wav} and~\ref{fig:bit32-wav} show the waveforms encrypted by shuffling and flipping, respectively, under the condition $M=10$.
Figures~\ref{fig:ori-spe}-\ref{fig:bit32-spe} correspond to the spectrograms in Figs.~\ref{fig:ori-wav}-\ref{fig:bit32-wav}, respectively.
By comparing the original and encrypted waveforms, we can see that there are changes in the outline, but they are minor.
By contrast, from Figs.~\ref{fig:pix32-spe} and \ref{fig:bit32-spe}, it can be seen that frequency the response of the original waveform has been significantly changed by the encryption.
These characteristics led to performance degradation without the correct key.


\subsection{Evaluation of privacy-preserving performance}
\begin{figure}[t]
\centering
\begin{minipage}[b]{0.3\linewidth}
    \centering
    \includegraphics[keepaspectratio,width=2cm]{fig/ori-spe.pdf}
    \subcaption{Original}
    \label{fig:ori3}
  \end{minipage}
  \hspace{0.01\columnwidth}
  \begin{minipage}[b]{0.3\linewidth}
    \centering
    \includegraphics[keepaspectratio, width=2cm]{fig/pix3-spe.pdf}
    \subcaption{Encryption}
    \label{fig:pix8-2}
  \end{minipage}
  \hspace{0.01\columnwidth}
  \begin{minipage}[b]{0.3\columnwidth}
    \centering
    \includegraphics[keepaspectratio, width=2cm]{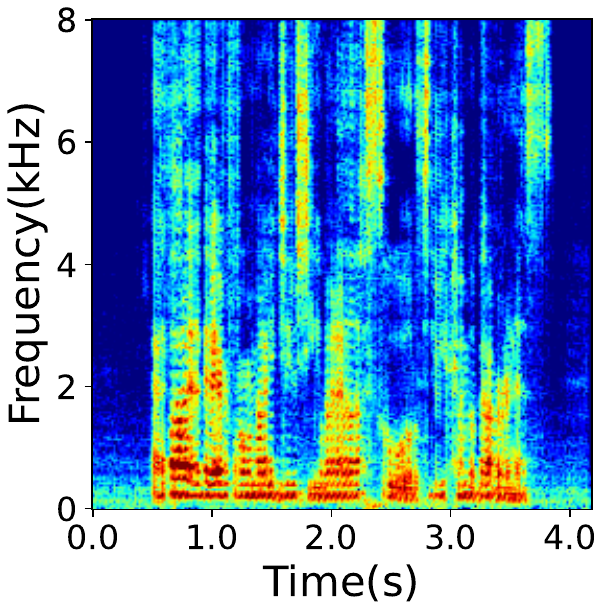}
    \subcaption{Reconstruction}
    \label{fig:pix-pghi8}
  \end{minipage}
  \caption{Examples of phase reconstruction of the spectrogram encrypted by shuffling ($M=3$)}
  \label{figs:pghi-pix}
\end{figure}

\begin{figure}[t]
\centering
  \begin{minipage}[b]{0.3\linewidth}
    \centering
    \includegraphics[keepaspectratio,width=2cm]{fig/ori-spe.pdf}
    \subcaption{Original}
    \label{fig:ori4}
  \end{minipage}
  \hspace{0.01\columnwidth}
  \begin{minipage}[b]{0.3\linewidth}
    \centering
    \includegraphics[keepaspectratio, width=2cm]{fig/bit3-spe.pdf}
    \subcaption{Encryption}
    \label{fig:bit8-2}
  \end{minipage}
  \hspace{0.01\columnwidth}
  \begin{minipage}[b]{0.3\columnwidth}
    \centering
    \includegraphics[keepaspectratio, width=2cm]{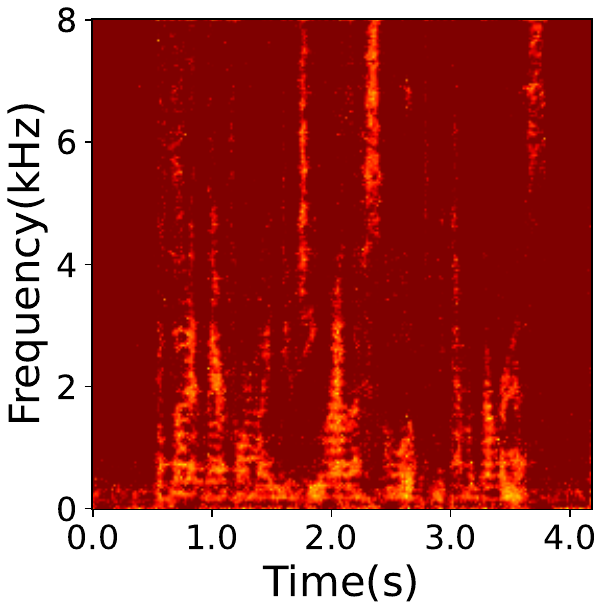}
    \subcaption{Reconstruction}
    \label{fig:bit-pghi8}
  \end{minipage}
  \caption{Examples of phase reconstruction of the spectrogram encrypted by flipping ($M=3$)}
  \label{figs:pghi-bit}
\end{figure}

To evaluate the security of encryption methods, phase reconstruction was performed using Průša \textit{et al.}'s method.
Figure~\ref{figs:pghi-pix} shows the results of the phase reconstruction of the spectrogram encrypted by shuffling.
We first applied shuffling to the spectrogram of the original speech (Fig.~\ref{fig:ori3}) under $M=3$. Then, we reconstructed the speech by applying phase reconstruction to the encrypted spectrogram (Fig.~\ref{fig:pix8-2}). The spectrogram of the reconstructed speech is shown in Fig.~\ref{fig:pix-pghi8}.
Figures~\ref{fig:ori3} and~\ref{fig:pix-pghi8} show that the structure of the spectrogram of the original speech and that of the speech obtained by phase reconstruction were different.
The original speech and the speech obtained by phase reconstruction were also different.
Figure~\ref{figs:pghi-bit} shows the results of the phase reconstruction of the spectrogram encrypted by flipping.
Spectrograms in Fig.~\ref{figs:pghi-bit} were obtained by the same procedure as those for shuffling.
Figures~\ref{fig:ori4} and~\ref{fig:bit-pghi8} show that the overall structure and amplitude values of the two spectrograms are significantly different.
The original speech was hardly audible from the speech obtained by phase reconstruction.
From the results in Figs.~\ref{figs:pghi-pix} and~\ref{figs:pghi-bit}, we found that it is difficult to reconstruct the original speech from the spectrogram encrypted by the proposed methods.

\begin{figure}[t]
\centering
\begin{minipage}[b]{0.3\linewidth}
    \centering
    \includegraphics[keepaspectratio,width=2cm]{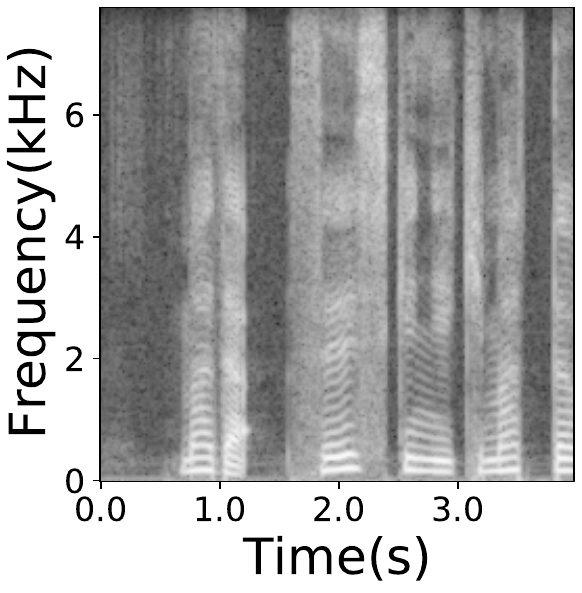}
    \subcaption{Original}
    \label{fig:ori-attack1}
  \end{minipage}
  \hspace{0.01\columnwidth}
  \begin{minipage}[b]{0.3\linewidth}
    \centering
    \includegraphics[keepaspectratio, width=2cm]{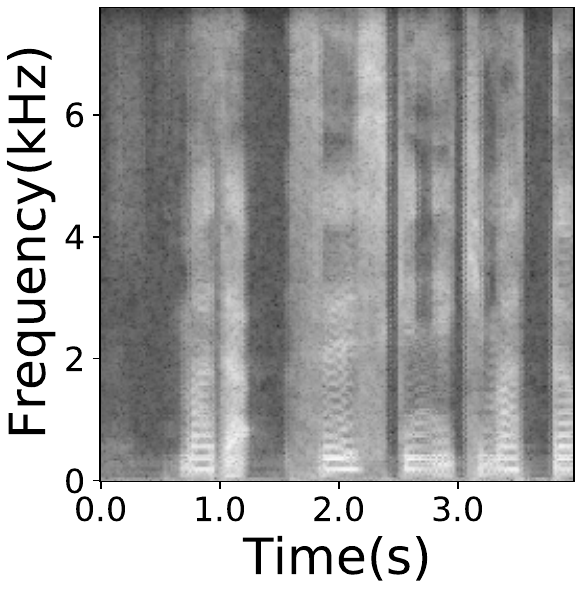}
    \subcaption{Encrypted}
    \label{fig:pix-enc}
  \end{minipage}
  \hspace{0.01\columnwidth}
  \begin{minipage}[b]{0.3\columnwidth}
    \centering
    \includegraphics[keepaspectratio, width=2cm]{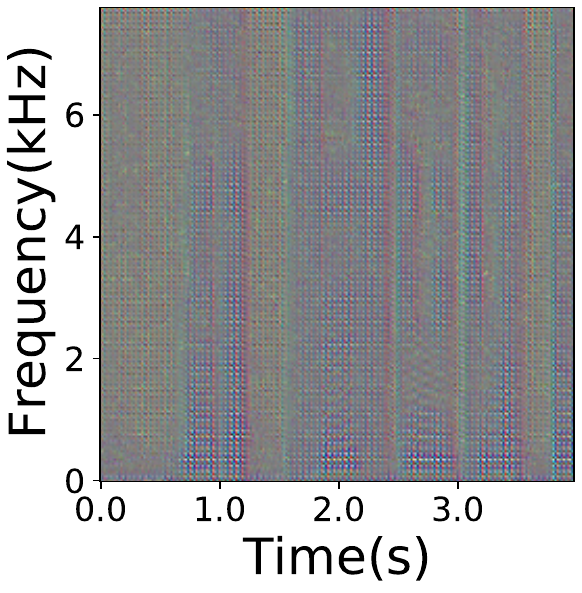}
    \subcaption{Decrypted}
    \label{fig:pix-dec}
  \end{minipage}
  \caption{Examples of decryption of the spectrogram encrypted by shuffling ($M=3$)}
  \label{figs:attack-pix}
\end{figure}

\begin{figure}[t]
\centering
  \begin{minipage}[b]{0.3\linewidth}
    \centering
    \includegraphics[keepaspectratio,width=2cm]{fig/img_clip-ori.pdf}
    \subcaption{Original}
    \label{fig:ori-attack2}
  \end{minipage}
  \hspace{0.01\columnwidth}
  \begin{minipage}[b]{0.3\linewidth}
    \centering
    \includegraphics[keepaspectratio, width=2cm]{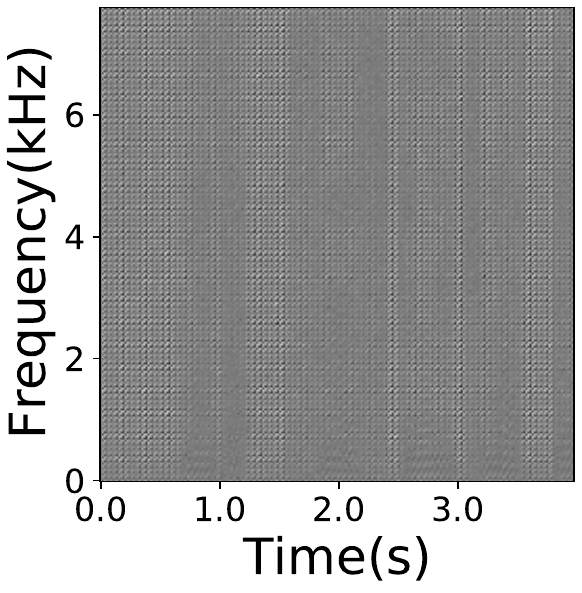}
    \subcaption{Encrypted}
    \label{fig:bit-enc}
  \end{minipage}
  \hspace{0.01\columnwidth}
  \begin{minipage}[b]{0.3\columnwidth}
    \centering
    \includegraphics[keepaspectratio, width=2cm]{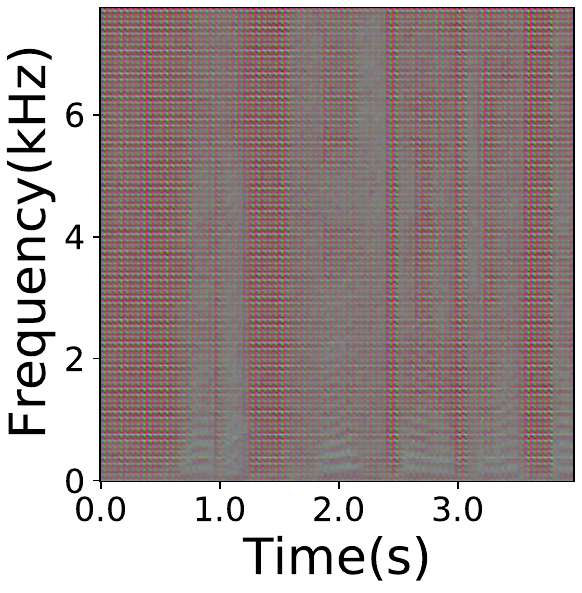}
    \subcaption{Decrypted}
    \label{fig:bit-dec}
  \end{minipage}
  \caption{Examples of decryption of the spectrogram encrypted by flipping ($M=3$)}
  \label{figs:attack-bit}
  \vspace{-1mm}
\end{figure}

Then, a decryption attack was performed on the encrypted spectrograms using Alex \textit{et al.}'s method.
The block size must be known when the attacker decrypts the encrypted data.
Therefore, in the experiments, we assumed that the attacker knows the block size.
Generally, spectrograms can be treated as grayscale images, but since Alex \textit{et al.}' s method only supports 8-bit RGB images, we scale the spectrogram so that the maximum value is $255$ and the minimum value is $0$.
Figure~\ref{figs:attack-pix} shows the result of attacking a spectrogram encrypted by shuffling under the condition $M=3$, and Fig.~\ref{figs:attack-bit} shows the result of attacking a spectrogram encrypted by flipping under the condition $M=3$.
From Figs.~\ref{fig:pix-dec} and~\ref{fig:bit-dec}, we found that spectrograms could not be completely decrypted even if the attacker attacked with the existing method.
Note also that this approach requires knowing the block size in advance. 
In this experiment, $M$ was set to a small value to match the existing model structure.
However, a larger $M$ makes the proposed method even more robust because the key space is larger and decryption becomes more difficult.

\section{Conclusion}\label{seq:conclude}
In this paper, we proposed the privacy-preserving methods using a secret key for CNN-based models: shuffling and flipping.
The encrypted spectrograms and waveforms obtained by the proposed methods were difficult to use without the correct key in the ASR and ASV tasks. 
In addition, to evaluate the privacy-preserving performance of the proposed encryption method, the phase reconstruction and decryption attack methods were applied to the encrypted data.
Our experiments showed that only the authorized user who knows the correct key could use the classification system correctly. 
The robustness of the proposed methods against existing attack methods was also confirmed.
As future work, we will develop a novel spectrogram and waveform encryption method that is less sensitive to block size and also further evaluate the robustness of the proposed method against existing decryption methods.

\section{Acknowledgements}
This work was supported in part by SECOM Science and Technology Foundation.

\bibliographystyle{IEEEtran}
\bibliography{IEEEabrv,refs}

\end{document}